\pdfoutput=1

\documentclass[11pt,twoside,a4paper,cmspaper,final,collab]{cms-tdr}
\def\svnVersion{454045P}\def\cmsCernNoTag{CERN-EP-2017-205}\def\cmsCernDate{\today}\def\cmsMessage{Published in Physical Review Letters as \href{http://dx.doi.org/10.1103/PhysRevLett.120.142302}{\doi{10.1103/PhysRevLett.120.142302.}}}

\begin{document}\cmsNoteHeader{HIN-16-006}

\hyphenation{had-ron-i-za-tion}
\hyphenation{cal-or-i-me-ter}
\hyphenation{de-vices}
\RCS$Revision: 445376 $
\RCS$HeadURL: svn+ssh://svn.cern.ch/reps/tdr2/papers/HIN-16-006/trunk/HIN-16-006.tex $
\RCS$Id: HIN-16-006.tex 445376 2018-02-11 21:51:56Z alverson $

\providecommand{\sqrtsNN}{\ensuremath{\sqrt{\smash[b]{s_{_{\mathrm{NN}}}}}}\xspace}
\newcommand {\PbPb}  {\ensuremath{\mathrm{PbPb}}\xspace}
\newcommand {\pPb}  {\ensuremath{\Pp\mathrm{Pb}}\xspace}
\newcommand{\Pb}{\ensuremath{\mathrm{Pb}}\xspace}
\newcommand{\pp}{\ensuremath{\Pp\Pp}\xspace}

\providecommand{\HYDJET} {{\textsc{hydjet}}\xspace}
\newcommand{\JEWEL} {{\textsc{jewel}}\xspace}
\providecommand{\ptjet}{\ensuremath{p_{\mathrm{T,jet}}}\xspace}
\providecommand{\zg}{\ensuremath{z_{\mathrm{g}}}\xspace}
\newcommand{\mylumipp}{\ensuremath{ 27.4\pbinv}\xspace}
\newcommand{\mylumipbpb}{\ensuremath{ 404\mubinv}\xspace}

\cmsNoteHeader{HIN-16-006}
\title{\texorpdfstring{Measurement of the splitting function in \pp and \PbPb collisions at $\sqrtsNN=5.02\TeV$}{Measurement of the splitting function in pp and PbPb collisions at sqrt(s[NN])=5.02 GeV}}

\date{\today}

\abstract{
Data from heavy ion collisions suggest that the evolution of a parton shower is modified by interactions with the color charges in the dense partonic medium created in these collisions, but it is not known where
in the shower evolution the modifications occur.
The momentum ratio of the two leading partons, resolved as subjets,
provides information about the parton shower evolution.
This substructure observable, known as the splitting function, reflects the process of a parton splitting into two other partons and has been measured for jets with transverse momentum between 140 and 500\GeV,
in \pp and \PbPb collisions at a center-of-mass energy of 5.02\TeV per nucleon pair.
In central \PbPb collisions, the splitting function indicates a more unbalanced momentum ratio,
compared to peripheral \PbPb and \pp collisions.
The measurements are compared to various predictions from event generators and analytical calculations.}

\hypersetup{%
pdfauthor={CMS Collaboration},%
pdftitle={Measurement of the splitting function in pp and PbPb collisions at sqrt(s[NN])=5.02 GeV},%
pdfsubject={CMS},%
pdfkeywords={CMS, heavy ions, jet quenching, splitting function}}

\maketitle

Scattering processes with large momentum transfer $Q$ between the partonic constituents of colliding nucleons occur early in heavy ion collisions.
Further interactions of the outgoing partons with the produced
(colored) hot and dense quantum chromodynamics (QCD) medium
(the quark-gluon plasma, QGP)
may modify the angular and momentum distributions of final-state hadronic jet fragments relative to those in proton-proton collisions.
This process, known as jet quenching, can be used to probe the properties of the QGP~\cite{Gyulassy:1990ye,Baier:1994bd}.
Jet quenching was first observed at RHIC~\cite{Adcox:2001jp,Adler:2003qi,Adams:2003kv,Arsene:2003yk,Back:2003qr,Adamczyk:2016fqm,Adamczyk:2017yhe}
and then at the LHC~\cite{Aamodt:2010jd,Aamodt:2011vg,Aad:2015wga,CMS:2012aa,Aad:2010bu,Chatrchyan:2012nia,Aad:2012vca,Abelev:2013kqa,Adam:2015ewa,Khachatryan:2016odn,Adam:2015doa,Chatrchyan:2013kwa,Chatrchyan:2014ava,Aaboud:2017bzv,Acharya:2017goa,Khachatryan:2015lha}.
This Letter reports an attempt to isolate parton splittings to two well separated partons with high transverse momentum (\pt),
probing medium induced effects during the parton shower evolution in the QGP.
Information about these leading partons of a hard splitting can be obtained by removing the softer wide-angle radiation contributions, done through the use of jet grooming algorithms that attempt to split (``decluster'') a single jet into two subjets~\cite{Ellis:2009me,Butterworth:2008iy,Krohn:2009th,Dasgupta:2013ihk,Larkoski:2014wba}.
For a parton shower in vacuum, these subjets provide access to the properties of the first splitting in the parton evolution~\cite{Altarelli:1977zs,Larkoski:2015lea}. Interactions of the two outgoing partons with the QGP potentially modify the properties of subsequent splittings resulting in different subjet properties.
This Letter reports a study of hard parton splittings in \pp and \PbPb collisions.

An observable characterizing the parton splitting, denoted by \zg, is defined as the ratio between the \pt of the less energetic subjet, $p_{\mathrm{T},2}$, and the \pt sum of the two subjets~\cite{Larkoski:2015lea}, \mbox{$\zg = {p_{\mathrm{T},2}} / ({p_{\mathrm{T},1}+p_{\mathrm{T},2}})$}.
A measurement of the \zg distribution in \pp collisions,
using CMS open data, was recently reported~\cite{Larkoski:2017bvj,Tripathee:2017ybi}.
In \PbPb collisions, this measurement reflects how the two color-charged partons produced in
the first splitting propagate through the QGP, probing the role of color coherence of the jet in the medium~\cite{Mehtar-Tani:2014yea}.
If the partons act as a single coherent emitter,
the two subjets will be equally modified,
leaving \zg unaffected~\cite{CasalderreySolana:2012ef}.
If, instead, the partons in the medium act as decoherent emitters,
the two subjets should be modified differently, thereby altering \zg.
In addition, \zg is sensitive to semi-hard medium-induced gluon radiation~\cite{Mehtar-Tani:2016aco},  modifications of the initial parton splitting~\cite{Chien:2016led}, and the medium response~\cite{Milhano:2017nzm}.

The analysis uses data collected by the CMS experiment in 2015.
The \PbPb and \pp data samples, both at a nucleon-nucleon center-of-mass energy of 5.02\TeV, correspond to integrated luminosities of $\mylumipbpb$ and $\mylumipp$, respectively.
The central feature of the CMS apparatus is a superconducting solenoid of 6\unit{m} internal diameter, providing a magnetic field of 3.8\unit{T}. Within the solenoid volume are a silicon pixel and strip tracker, a lead tungstate crystal electromagnetic calorimeter, and a brass and scintillator hadron calorimeter, each composed of a barrel and two endcap sections. Forward calorimeters extend the pseudorapidity, $\eta$, coverage provided by the barrel and endcap detectors.
A more detailed description of the CMS detector, together with a definition of the coordinate system used and the relevant kinematic variables, can be found in Ref.~\cite{Chatrchyan:2008zzk}.

The particle-flow (PF) algorithm reconstructs and identifies each individual particle with an optimized combination of information from the various elements of the CMS detector~\cite{Sirunyan:2017ulk}.
The PF candidates identified as a photon or a neutral hadron are treated as massless,
while for charged hadrons the pion mass is assumed.
The electron and muon PF candidates are assigned the corresponding lepton masses.
Jets are reconstructed from the PF candidates using the anti-\kt jet algorithm~\cite{Cacciari2011, Cacciari2006, Cacciari:2008gp} with a distance parameter $R=0.4$.
The kinematics of the jet are determined using the vectorial sum of all particle momenta in the jet.
For this analysis, jets are required to have $\ptjet > 140$\GeV and $\abs{\eta}< 1.3$.

The online event selection trigger also uses the anti-\kt algorithm with $R=0.4$ but applies a lower threshold on \ptjet:
all events with a PF jet with $\ptjet>80\GeV$ were recorded in the \pp case,
while in \PbPb collisions the triggers
(based on jets reconstructed from calorimeter deposits including a subtraction for the uncorrelated underlying event)
use a 100\GeV threshold.
Non-collision events, such as beam-gas interactions or cosmic-ray muons, are rejected offline~\cite{Khachatryan:2016odn}. The events are required to have a primary vertex reconstructed within 15\unit{cm} (0.15\unit{cm})
of the nominal interaction point along the beam direction (in the transverse plane).
The average number of additional collisions per bunch crossing is less than 0.9 in both data sets,
having a negligible effect on the measurement.
The \PbPb event sample is divided into centrality intervals, reflecting the impact parameter of the colliding nuclei,
using the percentage of the total inelastic hadronic cross section,
which is evaluated using the sum of the total energy deposited in both forward hadron calorimeters,
covering the $3<\abs{\eta}<5$ range~\cite{Chatrchyan:2011sx}.

The \PYTHIA6.423~\cite{Sjostrand:2006za} event generator (tune Z2*~\cite{Chatrchyan:2013gfi,Khachatryan:2015pea})
is used to calculate Monte Carlo (MC) corrections.
For \PbPb simulations, the \PYTHIA6 events are embedded into an underlying event produced with \HYDJET1.9~\cite{Lokhtin:2005px}.
All generated events undergo a full \GEANTfour~\cite{geant} simulation of the CMS detector response. Additional cross check samples are produced with \PYTHIA8.212~\cite{Sjostrand:2007gs} (tune CUETP8M1~\cite{Khachatryan:2015pea}) and {\HERWIGpp}~\cite{Bahr:2008pv} (tune EE5C~\cite{Seymour:2013qka}).

In \PbPb collisions, the constituents of the jet are corrected for the underlying event contribution using
the ``Constituent Subtraction'' method~\cite{Berta:2014eza},
a particle-by-particle approach that removes or corrects jet constituents based on the average underlying event density. The subtraction corrects both the four-momentum of the jet and its substructure.
Underlying event densities are determined by calculating the median \pt per unit area, $\rho$, and a density term related to the jet mass, $\rho_{\mathrm{m}}$, using a procedure in which all of the particles in the event are clustered into jets using the \kt algorithm with $R=0.4$~\cite{Soyez:2012hv,Cacciari2011, Cacciari2006}. To match the jets used in this analysis, only \kt jets with $\abs{\eta}<1.3$ are included in the density determination.
The influence of true hard jet fragments on the background estimation is reduced by excluding the two leading \kt jets.
The constituent-subtracted jets are corrected for the detector response with jet energy corrections derived from independent \pp and \PbPb simulations.
Additional corrections for the mismodeling of the detector response are also applied~\cite{Chatrchyan:2011ds}.

Jet grooming algorithms aim to isolate the hard prongs of a jet and remove soft wide-angle radiation.
The ``soft drop'' declustering procedure, used in this analysis,
is an extension of the modified mass drop tagger~\cite{Dasgupta:2013ihk}.
The procedure starts by selecting an anti-\kt{} jet that has already been constituent-subtracted and reclustered with the Cambridge/Aachen algorithm~\cite{Dokshitzer:1997in} to form a pairwise clustering tree with an angular-ordered structure. A pairwise declustering is performed on this tree. In each step of the declustering, a branching into two subjets is
accepted if they pass the soft drop condition~\cite{Larkoski:2014wba},
\begin{linenomath}
\begin{equation}
\frac{\mathrm{min}(p_{\mathrm{T},i},p_{\mathrm{T},j})}{p_{\mathrm{T},i}+p_{\mathrm{T},j}} > z_{\text{cut}}\left( \frac{\Delta R_{ij}}{R_{0}} \right)^{\beta},
\end{equation}
\end{linenomath}
where the subscripts ``$i$'' and ``$j$'' indicate the subjets at that step of the declustering, $\Delta R_{ij}$ is the distance between the two subjets in the $\eta$-$\phi$ plane, $R_{0}$ is the cone size of the anti-\kt{} jet, and $z_{\text{cut}}$ is an adjustable parameter.  If the soft drop condition is not satisfied, the softer subjet is dropped.
For this study, $z_{\text{cut}}$ is set to 0.1~\cite{Larkoski:2014wba}.
The parameter $\beta$ is set to 0,
which satisfies an extended version of infrared and collinear safety by absorbing the collinear divergences into a generalized fragmentation function recovering the QCD splitting function~\cite{Larkoski:2015lea}.
Once the soft drop condition is satisfied, the two subjets at that position in the tree are used in the analysis.
If the soft drop condition is never satisfied, the jet is not used.
This is the case for 1.5\% of the jets measured at $\ptjet=140$\GeV,
increasing to 3.0\% at $\ptjet=300\GeV$, independent of collision centrality.

Groomed jets with a small distance between the two subjets frequently result from the ambiguous case where the two subjets cannot be distinctly resolved,
leading to a significant misassignment of particle constituents to subjets.
An additional selection of $\Delta R_{12} > 0.1$ is applied,
removing 40\% (60\%) of the jets measured at low (high) \ptjet,
to avoid an unphysical modification of \zg.
This selection rejects an additional 15\% (5\%) of the jets at low (high) \ptjet in the 10\% most central PbPb collisions,
in comparison to the non-central collisions,
an effect well reproduced by the simulation.
The systematic uncertainty on the \zg variable
is evaluated by varying the $\Delta R_{12}$ minimum distance
requirement by its one standard deviation MC resolution of 10\%;
this variation results in a 2\% uncertainty, independent of centrality.

\begin{figure}[h!t]
\centering
\includegraphics[width=0.5\textwidth]{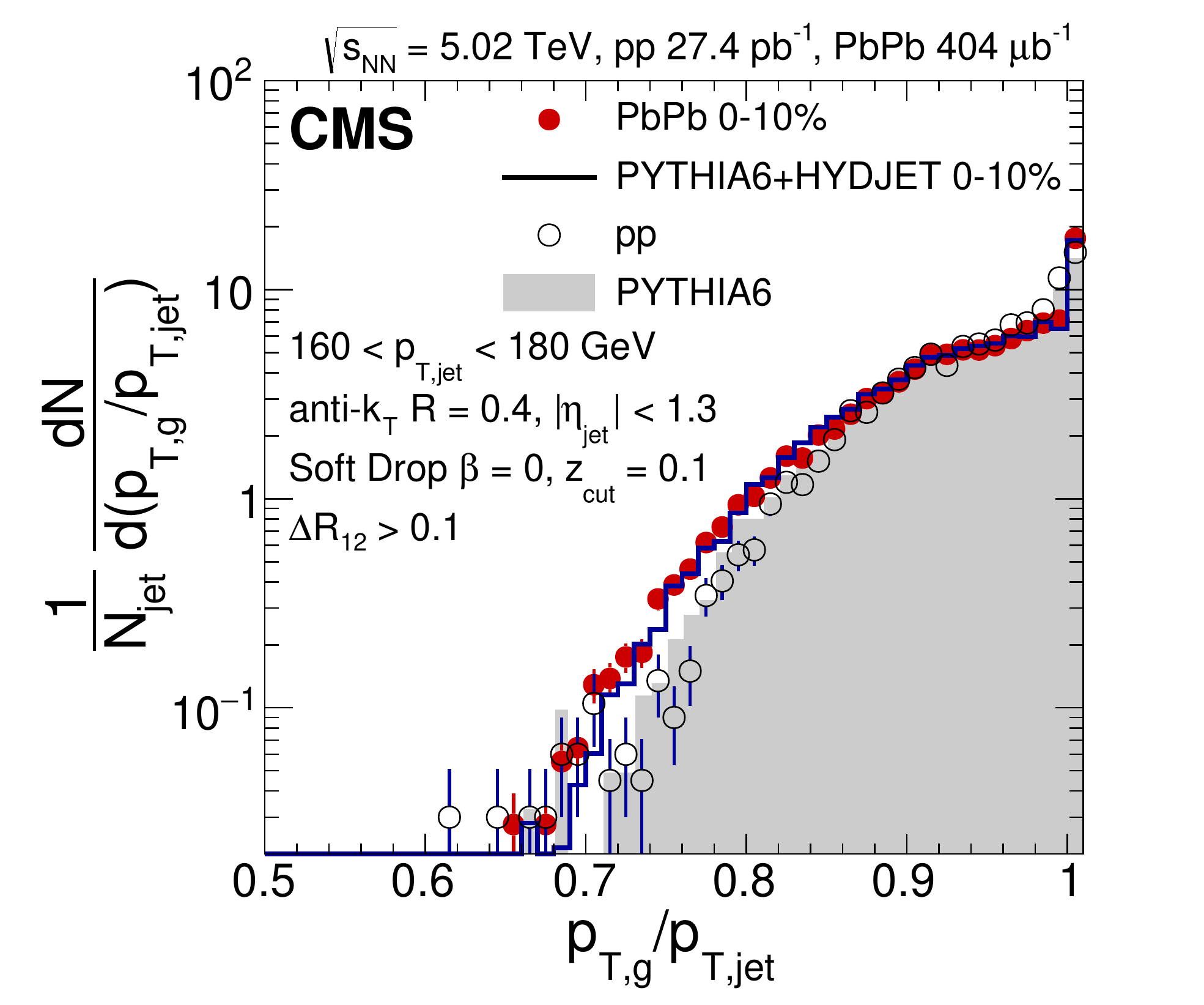}
       \caption{Groomed jet energy fraction in \pp and in the 10\% most central \PbPb collisions,
       for jets with $160<\ptjet<180\GeV$ and $|\eta_{\text{jet}}|<1.3$.
       The \pp (\PbPb) data are compared to \mbox{\PYTHIA6} (embedded in \HYDJET) distributions.}
        \label{fig:GroomedEnergy}
\end{figure}

The transverse momentum of the jet after grooming, $p_{\mathrm{T,g}}$, is identical to or smaller than the original \ptjet.
The groomed \pt fraction, $p_{\mathrm{T,g}}/\ptjet$,
is compared to simulations in Fig.~\ref{fig:GroomedEnergy} for jets with $160<\ptjet<180\GeV$,
in $\Pp\Pp$ and central \PbPb collisions.
The measured and simulated distributions are in agreement.

The potential bias due to the online jet trigger is evaluated by using events collected with a lower threshold and also minimum bias events.
For the 10\% most central \PbPb collisions, a bias is found in the lowest \ptjet range, $140<\ptjet<160\GeV$,
changing the yield by values linearly decreasing from $+6$\% at $\zg = 0.1$ to $-15$\% at $\zg = 0.5$.
In the 10--30\% centrality class, the bias is half as large, and it vanishes for more peripheral events.
The full bias is corrected for and the magnitude of the correction is treated as a \zg systematic uncertainty.
The trigger has no effect on the measurements at higher \ptjet.

The systematic uncertainty in the jet energy scale,
on the measured and simulated distributions,
is obtained by propagating the uncertainties in the jet response correction~\cite{Chatrchyan:2011ds,Khachatryan:2016kdb}.
A maximum deviation in yield of 4\% is found in central \PbPb collisions, decreasing to 2\% in \pp and peripheral \PbPb collisions.
This effect tends to increase (decrease) the \pt{} of the leading (subleading) subjet.
The systematic uncertainty in the normalization of the \zg distributions is estimated to be 5\%\,(3\%) in central\,(peripheral) collisions. The relative uncertainty in the jet energy resolution is 10\%, leading to an uncertainty smaller than 0.5\% on the \zg distribution.

\begin{figure}[h!t]
\centering
       \includegraphics[width=0.5\textwidth]{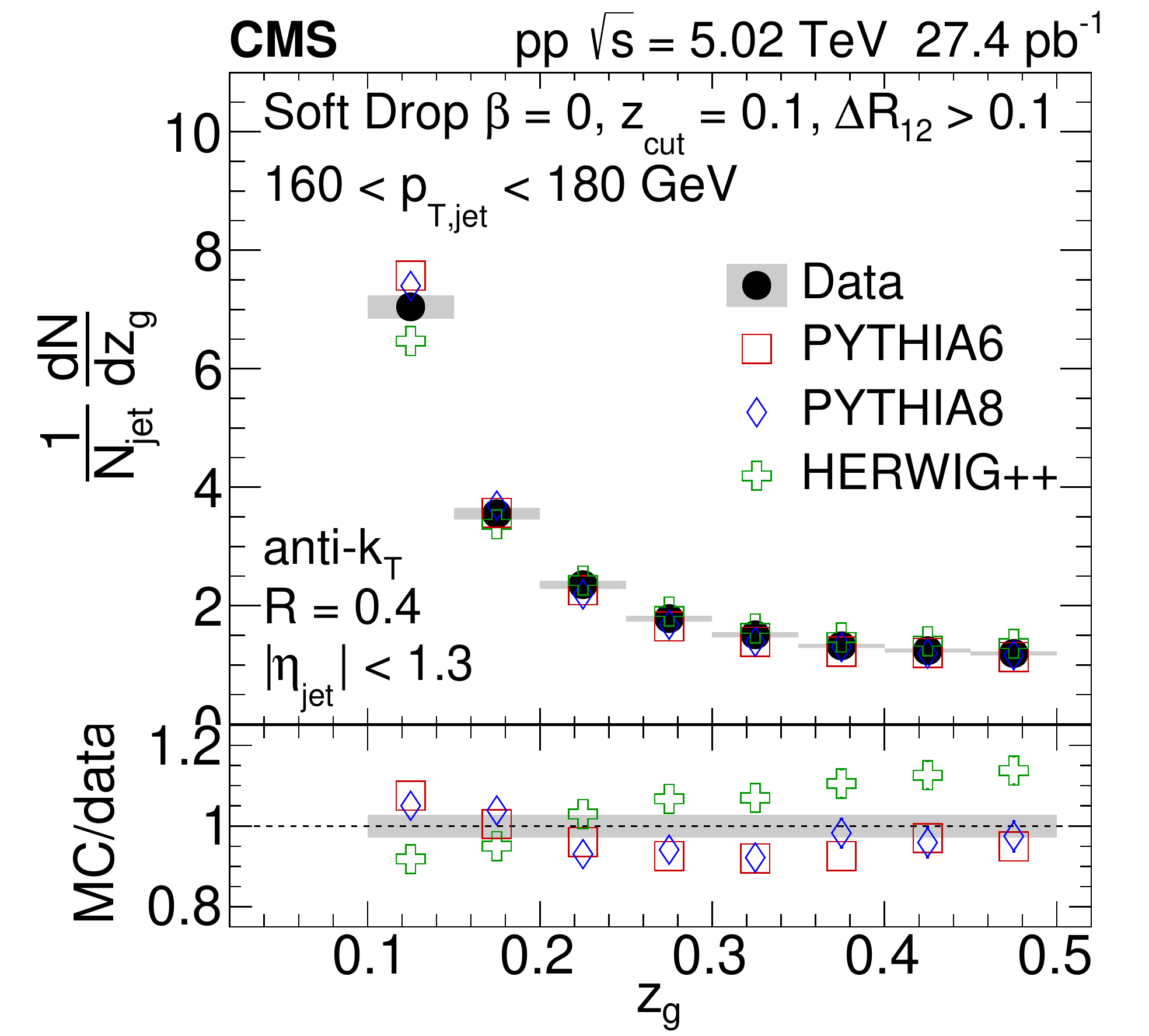}
       \caption{The \zg distribution in \pp collisions for $160<p_{\mathrm{T,jet}}<180\GeV$,
       compared to predictions from event generators.
       The error bars (shaded area) represent the statistical (systematic) uncertainty.}
       \label{fig:ZgPPPt4}
\end{figure}

Figure~\ref{fig:ZgPPPt4} shows the \zg distribution measured in \pp collisions,
together with results obtained with \PYTHIA6, \PYTHIA8, and \HERWIGpp,
including a full simulation of detector effects.
Both \PYTHIA simulations have a slightly steeper \zg distribution than the data,
while \HERWIGpp shows an opposite trend.

To compare the \zg distribution in \pp and \PbPb collisions, in given \ptjet and centrality ranges,
the measurements in \pp collisions are adjusted to match the subjet resolution in \PbPb data.
The resolution correction is derived, for each \ptjet and collision centrality range,
from full detector simulation studies of the ratio of the \zg distributions
between \PYTHIA and \PYTHIA embedded into \HYDJET.
The ratio between simulated \PbPb and \pp \zg distributions shows a relative decrease in the number of \PbPb events at high \zg,
reaching $\sim$40\% in central collisions and negligible in peripheral collisions.
The uncertainty in the correlation between the response of the two subjets is estimated by varying the individual subjet resolution by 10\%, the relative correlation by 15\%, and the subjet energy scale by 5\%, corresponding to one standard deviation in resolution.
This results in an uncertainty of 8--10\% in \zg.
The mismodeling of the \zg distribution in \PYTHIA, evaluated by reweighting to the \zg measurement in \pp collisions,
adds an uncertainty of 4--5\%.
These uncertainties are assigned to the ``smeared'' \pp data points. The resolution correction is validated with a parametric resolution model that uses the jet resolution and a sampled \zg in each \ptjet range, and recreates the correction function for each centrality selection by sampling the individual subjet resolutions.

\begin{figure}[th]
  \centering
  \includegraphics[width=0.45\textwidth]{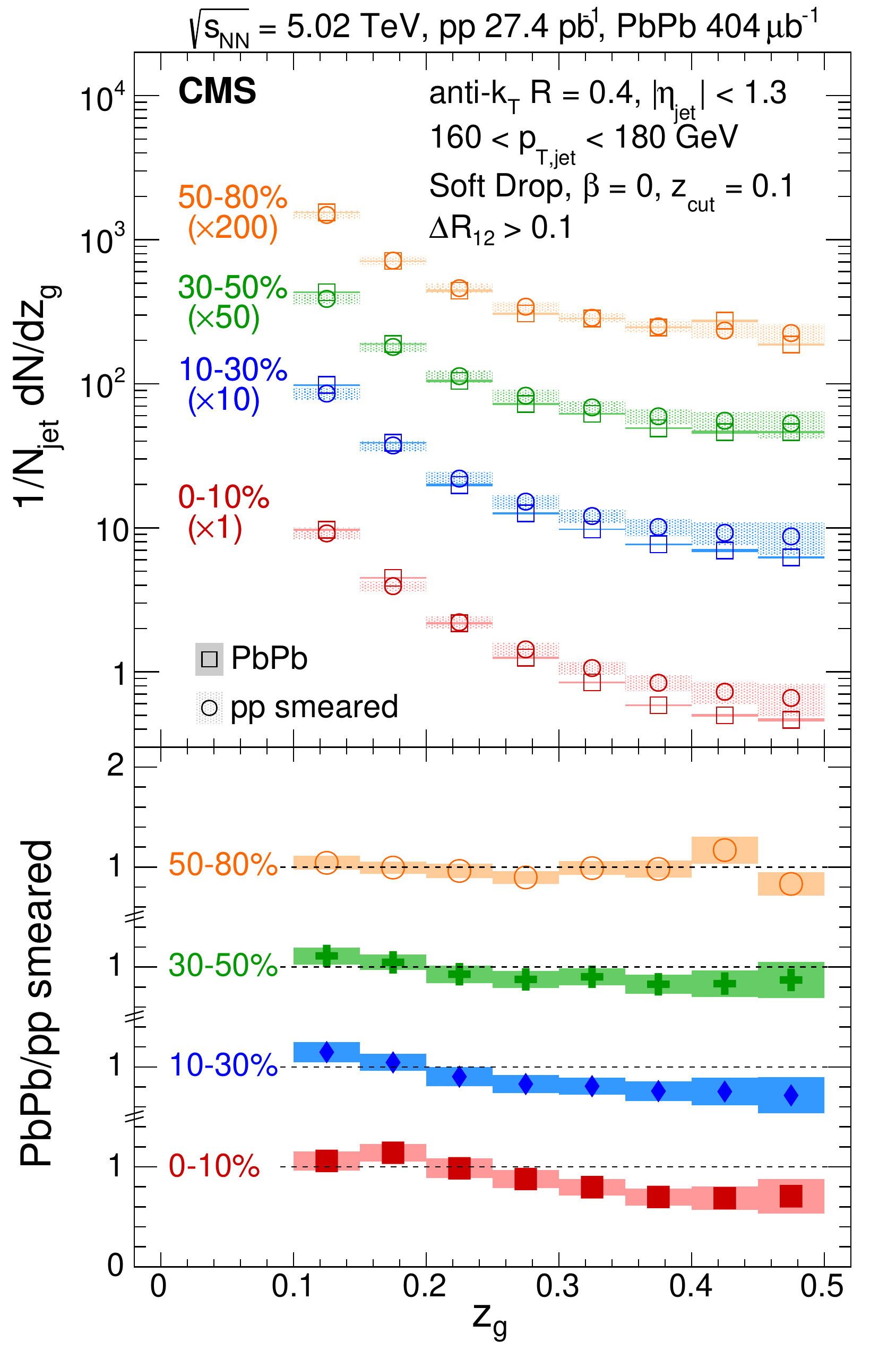}
  \caption{The \zg distributions in \PbPb collisions for $160<p_{\mathrm{T,jet}}<180\GeV$, in several centrality ranges, compared to \pp data smeared to account for the differences in resolution. The error bars (shaded area) represent the statistical (systematic) uncertainty.
  }
  \label{fig:ZgPbPbPPPtBin4}
\end{figure}
\begin{figure}[th]
  \centering
  \includegraphics[width=0.4\textwidth]{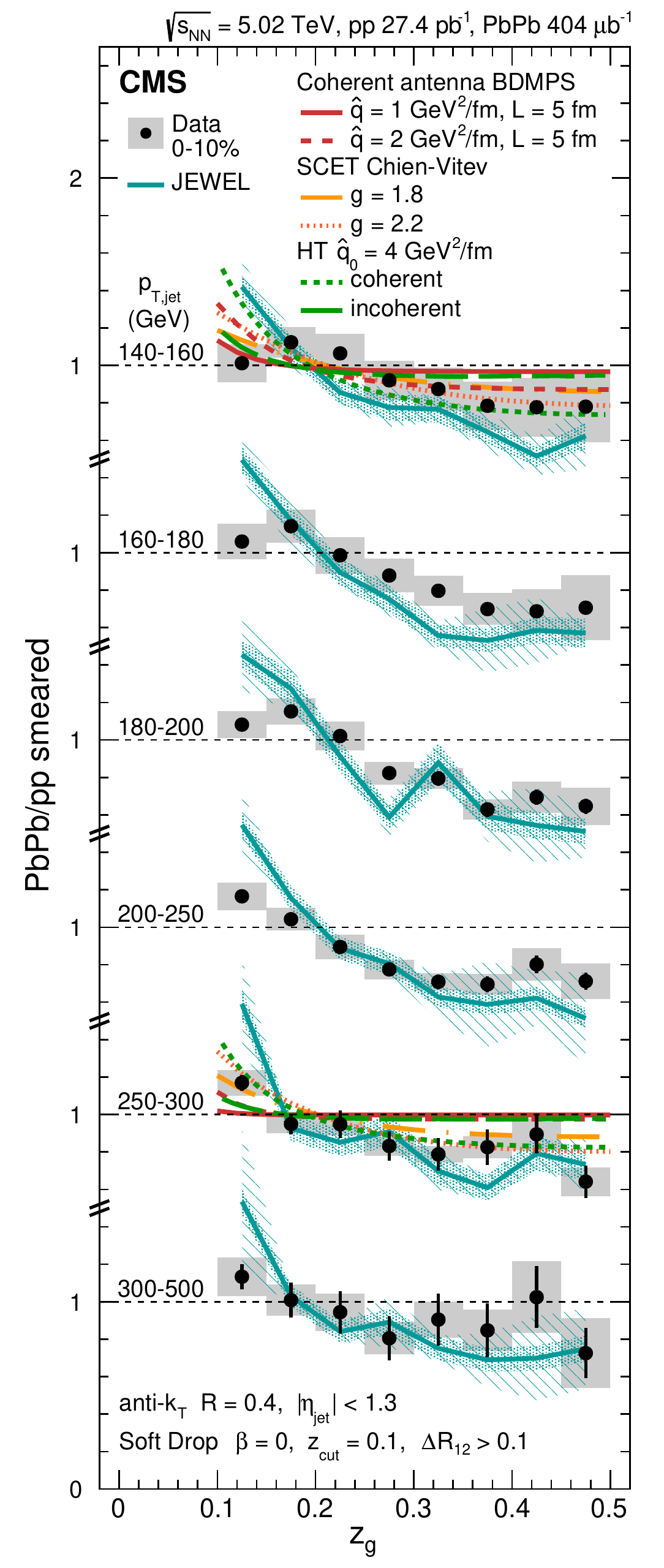}
  \caption{Ratios of \zg distributions in \PbPb and smeared \pp collisions in the 10\% most central events, for several \ptjet ranges, compared to various jet quenching theoretical calculations~\cite{Chien:2016led,Mehtar-Tani:2016aco,Chang:2017gkt,Milhano:2017nzm}. The error bars (shaded area) represent the statistical (systematic) uncertainty. The diagonally hatched band denotes the uncertainty from the treatment of the medium response using the \JEWEL event generator.
  }
  \label{fig:ZgRatioCentralPt}
\end{figure}
Figure~\ref{fig:ZgPbPbPPPtBin4} shows the \zg distributions measured in \PbPb collisions,
for several centrality intervals, in comparison to the smeared \pp reference data.
The systematic uncertainties on the \zg distributions are fully correlated from point-to-point, resulting in an anti-correlated uncertainty on the self-normalized distributions, and are uncorrelated between the \pp and \PbPb data sets.
The \zg distribution in peripheral \PbPb collisions agrees with the \pp reference,
while the more central collisions exhibit a steeper \zg distribution.
Differences between the \zg of quark- and gluon-initiated jets are found to be a few percent~\cite{Larkoski:2015lea},
so that the observed modification cannot be attributed to the flavor composition within a fixed \ptjet interval.
The observation indicates that the splitting into two branches becomes increasingly more unbalanced as the \PbPb collisions become more central.

The modification of the \zg distribution in central \PbPb collisions is shown in Fig.~\ref{fig:ZgRatioCentralPt} over a wide kinematic range in \ptjet.
The measurement is compared to a prediction of the \JEWEL event generator (shown with statistical and theoretical uncertainties originating from the treatment of the medium response), which incorporates medium-induced interactions while the partons propagate through the QGP~\cite{Zapp:2012ak,KunnawalkamElayavalli:2017hxo,Milhano:2017nzm}.
The measurement is also compared
with a soft-collinear effective theory (SCET) with Glauber gluon interactions~\cite{Chien:2016led} for two different quenching strengths,
with a calculation incorporating multiple medium-induced gluon bremsstrahlung (BDMPS)~\cite{Baier:1994bd,Baier:1996kr,Baier:1996sk} assuming that the two hard partons radiate gluons as a coherent emitter~\cite{Mehtar-Tani:2016aco},
and with a higher twist (HT) approach employing both coherent and incoherent energy loss~\cite{Chang:2017gkt}.
Each of the three models is presented for two settings of the parameters
reflecting their medium properties, as indicated in the legends,
where $L$ is the medium length, $\hat{q}$ and $\hat{q_{0}}$ denote medium transport coefficients,
and $g$ is the coupling strength between the jet and the medium.
The BDMPS medium effect is too weak to describe the observed \ptjet dependence,
while the other models reproduce the data at low and high \ptjet,
using medium properties previously tuned to match measurements
of the nuclear modification factors of charged hadrons and jets.
For the HT calculation, the presence or absence of color coherence makes a significant difference.
Since the detector resolution effects have a negligible impact on the theoretical calculations,
given that they largely cancel in the \PbPb to (smeared) \pp ratio,
the theoretical curves are shown without detector smearing.

In summary, the first measurement of the splitting function in \pp and \PbPb collisions at a center-of-mass energy of 5.02\TeV per nucleon pair has been presented.
This represents the first application of a grooming technique to \PbPb data,
removing soft wide-angle radiation from the jet and thereby isolating the two leading subjets.
The momentum sharing between these subjets is used to obtain information about hard parton splitting processes during the shower evolution.
The \PYTHIA and \HERWIGpp event generators reproduce the measured splitting function in \pp and peripheral \PbPb collisions,
at the level of 15\%.
In central \PbPb collisions, a steeper \zg distribution is observed,
indicating that the parton splitting process is modified by the hot medium created in heavy ion collisions.
These results provide new insight into the role of color coherence and other attributes of the interactions of partons in the quark-gluon plasma.

\begin{acknowledgments}
We congratulate our colleagues in the CERN accelerator departments for the excellent performance of the LHC and thank the technical and administrative staffs at CERN and at other CMS institutes for their contributions to the success of the CMS effort. In addition, we gratefully acknowledge the computing centers and personnel of the Worldwide LHC Computing Grid for delivering so effectively the computing infrastructure essential to our analyses. Finally, we acknowledge the enduring support for the construction and operation of the LHC and the CMS detector provided by the following funding agencies: BMWFW and FWF (Austria); FNRS and FWO (Belgium); CNPq, CAPES, FAPERJ, and FAPESP (Brazil); MES (Bulgaria); CERN; CAS, MoST, and NSFC (China); COLCIENCIAS (Colombia); MSES and CSF (Croatia); RPF (Cyprus); SENESCYT (Ecuador); MoER, ERC IUT, and ERDF (Estonia); Academy of Finland, MEC, and HIP (Finland); CEA and CNRS/IN2P3 (France); BMBF, DFG, and HGF (Germany); GSRT (Greece); OTKA and NIH (Hungary); DAE and DST (India); IPM (Iran); SFI (Ireland); INFN (Italy); MSIP and NRF (Republic of Korea); LAS (Lithuania); MOE and UM (Malaysia); BUAP, CINVESTAV, CONACYT, LNS, SEP, and UASLP-FAI (Mexico); MBIE (New Zealand); PAEC (Pakistan); MSHE and NSC (Poland); FCT (Portugal); JINR (Dubna); MON, RosAtom, RAS, RFBR and RAEP (Russia); MESTD (Serbia); SEIDI, CPAN, PCTI and FEDER (Spain); Swiss Funding Agencies (Switzerland); MST (Taipei); ThEPCenter, IPST, STAR, and NSTDA (Thailand); TUBITAK and TAEK (Turkey); NASU and SFFR (Ukraine); STFC (United Kingdom); DOE and NSF (USA). \end{acknowledgments}

\bibliography{auto_generated}

\cleardoublepage \appendix\section{The CMS Collaboration \label{app:collab}}\begin{sloppypar}\hyphenpenalty=5000\widowpenalty=500\clubpenalty=5000\textbf{Yerevan Physics Institute,  Yerevan,  Armenia}\\*[0pt]
A.M.~Sirunyan, A.~Tumasyan
\vskip\cmsinstskip
\textbf{Institut f\"{u}r Hochenergiephysik,  Wien,  Austria}\\*[0pt]
W.~Adam, F.~Ambrogi, E.~Asilar, T.~Bergauer, J.~Brandstetter, E.~Brondolin, M.~Dragicevic, J.~Er\"{o}, M.~Flechl, M.~Friedl, R.~Fr\"{u}hwirth\cmsAuthorMark{1}, V.M.~Ghete, J.~Grossmann, J.~Hrubec, M.~Jeitler\cmsAuthorMark{1}, A.~K\"{o}nig, N.~Krammer, I.~Kr\"{a}tschmer, D.~Liko, T.~Madlener, I.~Mikulec, E.~Pree, N.~Rad, H.~Rohringer, J.~Schieck\cmsAuthorMark{1}, R.~Sch\"{o}fbeck, M.~Spanring, D.~Spitzbart, W.~Waltenberger, J.~Wittmann, C.-E.~Wulz\cmsAuthorMark{1}, M.~Zarucki
% ...
\vskip\cmsinstskip
% ...
1:~~Also at Vienna University of Technology, Vienna, Austria\\
% ...

\end{sloppypar}
\end{document}